\documentclass[english,letterpaper,pra]{revtex4}
\usepackage[T1]{fontenc}
\usepackage[latin1]{inputenc}
\usepackage{geometry}
\usepackage{graphicx}
\usepackage{amssymb}

\makeatletter

\usepackage{babel}
\makeatother
\begin{document}

\title{Coherent, adiabatic and dissociation regimes in coupled atomic-molecular
Bose-Einstein condensates}

\author{Pascal Naidon$^{1}$}

\email{pascal.naidon@nist.gov}

\author{, Eite Tiesinga$^{1,2}$ and Paul Julienne$^{1,2}$}

\address{$^{1}$Atomic Physics Division, $^{2}$Joint Quantum Institute, National
Institute of Standards and Technology, 100 Bureau Drive Stop 8423,
Gaithersburg, Maryland 20899-8423, USA}

\begin{abstract}
We discuss the dynamics of a Bose-Einstein condensate of atoms which
is suddenly coupled to a condensate of molecules by an optical or
magnetic Feshbach resonance. Three limiting regimes are found and
can be understood from the transient dynamics occuring for each pair
of atoms. This transient dynamics can be summarised into a time-dependent
shift and broadening of the molecular state. A simple Gross-Pitaevski\u{\i}
picture including this shift and broadening is proposed to describe
the system in the three regimes. Finally, we suggest how to explore
these regimes experimentally.
\end{abstract}
\maketitle

\section{Introduction}

Optical and magnetic Feshbach resonances \cite{fedichev1996,Tiesinga1993}
occur when two colliding particles are resonant with a bound state
\cite{feshbach1958}. They have been extensively used to probe and
control the properties of ultracold atomic gases \cite{kohler:1311}.
In particular, they give the ability to tune the interaction between
atoms, making ultracold gases very useful to test many-body theories
for a wide range of interaction strengths, which is generally not
possible in other kinds of systems. These resonances also led to the
controlled association of atoms into diatomic molecules, a process
called photoassociation \cite{jones:483} in the case of an optical
Feshbach resonance. This association is a coherent process and when
applied to degenerate gases, it can result in the formation of a Bose-Einstein
condensate of molecules, also called molecular condensate \cite{donley2002,herbig2003}.

In this work, we discuss the dynamics of an atomic Bose-Einstein condensate
which is suddenly coupled to a molecular condensate through an optical
or magnetic Feshbach resonance. In the case of an optical resonance,
this can be achieved by turning on a laser on resonance between a
colliding pair of atoms and an excited bound state. In the case of
a magnetic resonance, this can be achieved by suddenly imposing a
magnetic field which brings a bound state in resonance with the colliding
pair. In both cases, the two-atom system can be effectively described
by two channels: the (open) scattering channel where atoms collide
through an interaction potential $U(\vec{r})$ and a (closed) molecular
channel where the atoms can bind through an interaction potential
$U_{m}(\vec{r})$. The vector $\vec{r}$ denotes the relative position
of the two atoms. When $\vec{r}$ becomes large, the potential $U(\vec{r})\to0$
, and the potential $U_{m}(\vec{r})$ goes to the internal energy
$E_{m}$. The two channels are coupled by some potential $W(\vec{r})$.
In the case of an optical resonance, the two channels correspond to
different electronic states and the coupling is due to the laser.
In the case of a magnetic resonance, the two channels correspond to
different hyperfine states and the coupling is due to hyperfine interaction.
With this simple picture of the resonance at the two-atom level, we
can write equations for a Bose-Einstein condensate.

\section{Equations for coupled atomic-molecular condensate}

A coupled atomic-molecular Bose-Einstein condensate can be described
by time-dependent equations for its cumulants. Truncating to first
order as prescribed in Ref. \cite{kohler2002}, one obtains the following
set of equations \cite{kohler2003,gasenzer2004a,naidon2005}:\begin{eqnarray}
i\hbar\frac{\partial\Psi}{\partial t}(\vec{x}) & = & H_{x}\Psi(\vec{x})+\int\Psi^{*}(\vec{y})\big(U(\vec{x}-\vec{y})\Phi(\vec{x},\vec{y})+W(\vec{x}-\vec{y})\Phi_{m}(\vec{x},\vec{y})\big)d^{3}\vec{y}\label{eq:OriginalSet}\\
i\hbar\frac{\partial}{\partial t}\Phi^{\prime}(\vec{x},\vec{y}) & = & (H_{x}+H_{y})\Phi^{\prime}(\vec{x},\vec{y})+U(\vec{x}-\vec{y})\Phi(\vec{x},\vec{y})+W(\vec{x}-\vec{y})\Phi_{m}(\vec{x},\vec{y})\nonumber \\
i\hbar\frac{\partial}{\partial t}\Phi_{m}(\vec{x},\vec{y}) & = & (H_{x}+H_{y}-i\gamma/2)\Phi_{m}(\vec{x},\vec{y})+U_{m}(\vec{x}-\vec{y})\Phi_{m}(\vec{x},\vec{y})+W(\vec{x}-\vec{y})\Phi(\vec{x},\vec{y})\nonumber \end{eqnarray}
where $\vec{x}$ and $\vec{y}$ are space coordinates, $t$ is the
time variable (for clarity, it is omitted but implicitly present as
an argument of all functions), $\hbar$ is the reduced Planck's constant,
$H_{x}=-\frac{\hbar^{2}\nabla_{x}^{2}}{2M}+V(x)$ is the one-body
Hamiltonian for a single atom ($M$ is the atom mass and $V$ some
external trapping potential), $\Psi$ is the condensate wave function,
$\Phi_{m}$ is the pair wave function in the closed channel, and $\Phi$
is the pair wave function in the open channel. The function $\Phi^{\prime}(\vec{x},\vec{y})=\Phi(\vec{x},\vec{y})-\Psi(\vec{x})\Psi(\vec{y})$
is the pair cumulant. It corresponds to the deviation of the pair
wave function from its noninteracting form, which is a product of
two condensate wave functions. We included a loss term $-i\gamma/2$
in the closed channel to account for possible decay by spontaneous
emission. 

Usually, only one molecular level $\varphi_{m}$ of the potential
$U_{m}$ is in resonance with the atomic condensate pairs. Thus, we
can write:\begin{equation}
\Phi_{m}(\vec{x},\vec{y},t)=\Psi_{m}(\vec{R},t)\varphi_{m}(\vec{r})\label{eq:DecompositionPhim}\end{equation}
where $\vec{R}=(\vec{x}+\vec{y})/2$ and $\vec{r}=\vec{y}-\vec{x}$
are the centre-of-mass and relative coordinates. $\Psi_{m}$ is a
one-body field corresponding to the molecular condensate wave function,
describing the motion of a molecule whose internal state $\varphi_{m}$
satisfies\[
\big(-\frac{\hbar^{2}\nabla_{r}^{2}}{M}+U_{m}(\vec{r})\big)\varphi_{m}(\vec{r})=(E_{m}-E_{b})\varphi_{m}(\vec{r})\]
where $E_{b}$ is the binding energy associated with the molecular
state $\varphi_{m}$. We choose the normalisation $\langle\varphi_{m}\vert\varphi_{m}\rangle=1$,
and call $\Delta=E_{m}-E_{b}$ the detuning from the resonance.

After some approximations detailed in the appendix, we find that $\Psi$
and $\Psi_{m}$ satisfy the closed set of equations:

\begin{eqnarray}
i\hbar\frac{\partial\Psi}{\partial t}(\vec{x}) & = & \Big(-\frac{\hbar^{2}\nabla_{x}^{2}}{2M}+V(\vec{x})+g\vert\Psi(\vec{x})\vert^{2}\Big)\Psi(\vec{x})+\Psi^{*}(x)\Big(1+g\alpha f(t)\Big)w\Psi_{m}(\vec{x})\label{eq:Shifted1}\\
i\hbar\frac{\partial\Psi_{m}}{\partial t}(\vec{x}) & = & \Big(-\frac{\hbar^{2}\nabla_{x}^{2}}{4M}+2V(\vec{x})+\Delta-\Delta^{\prime}-i\gamma/2+w^{2}\alpha f(t)\Big)\Psi_{m}(\vec{x})+w\Psi(\vec{x})^{2}\label{eq:Shifted2}\end{eqnarray}
where $g$ is the coupling constant for elastic collisions between
atoms, $w$ is the coupling constant for atom-molecule conversion,
and $\Delta^{\prime}$ is a light shift of the molecular level due
to the coupling. For three-dimensional systems, we have\begin{eqnarray*}
\alpha & = & \pi/2\\
f(t) & = & \frac{1-i}{2\hbar}\Big(\frac{M}{h}\Big)^{3/2}\frac{1}{\sqrt{t}}\end{eqnarray*}
The term $w^{2}\alpha f(t)$ can be interpreted as a time-dependent
shift and broadening of the molecular level. It is a time-dependent
version of the static light shift $\Delta^{\prime}$: when the interaction
is switched on, a large shift and broadening of the molecular level
appears during some transient regime and then eventually goes to the
static value $\Delta^{\prime}$. The reason for the broadening is
the loss of molecules into atom pairs. Indeed, during a certain time,
the coupling $w$ can dissociate the molecules into the pair continuum
instead of dissociating them back to the atomic condensate state,
a process called {}``rogue dissociation'' in Ref. \cite{javanainen2002}.

\section{Coherent, adiabatic and dissociation regimes}

We now investigate the different regimes that one can get from Eqs.~(\ref{eq:Shifted1}-\ref{eq:Shifted2}).
These regimes can be identified by the qualitatively different short-time
dynamics. At short times, we can neglect the motion of molecules in
the trap. We can also write $\frac{\partial}{\partial t}\Psi_{m}(\vec{x},t)\approx(\Psi_{m}(\vec{x},t)-\Psi_{m}(\vec{x},0))/t$.
Using this with the initial conditions (\ref{eq:InitialConditions})
in Eq.~(\ref{eq:Shifted2}), we can solve for $\Psi_{m}(\vec{x})$
and insert it into Eq.~(\ref{eq:Shifted1}). We obtain an effective
Gross-Pitaevski\u{\i} equation for the condensate wave function $\Psi$\begin{equation}
i\hbar\frac{\partial\Psi}{\partial t}(\vec{x})=\Big(-\frac{\hbar^{2}\nabla_{x}^{2}}{2M}+V(\vec{x})+\Big(g-w^{2}\frac{1+g\alpha f(t)}{\Delta-\Delta^{\prime}-i\gamma/2+w^{2}\alpha f(t)-\frac{i\hbar}{t}}\Big)\vert\Psi(\vec{x})\vert^{2}\Big)\Psi(\vec{x})\label{eq:GPeq}\end{equation}
In the local density approximation, this equation leads to the familiar
rate equation\begin{equation}
\dot{\rho}(\vec{x},t)=-K(t)\rho(\vec{x},t)^{2}\label{eq:RateEquation}\end{equation}
where $\rho(\vec{x},t)=\vert\Psi(\vec{x},t)\vert^{2}$ is the atomic
condensate density, and $K(t)$ is a time-dependent rate coefficient\begin{equation}
K(t)=\frac{2}{\hbar}\mbox{Im}\frac{w^{2}+gw^{2}\alpha f(t)}{\Delta-\Delta^{\prime}-i\gamma/2+w^{2}\alpha f(t)-\frac{i\hbar}{t}}\label{eq:Rate}\end{equation}

\begin{figure}
\hfill{}\includegraphics[scale=1.8]{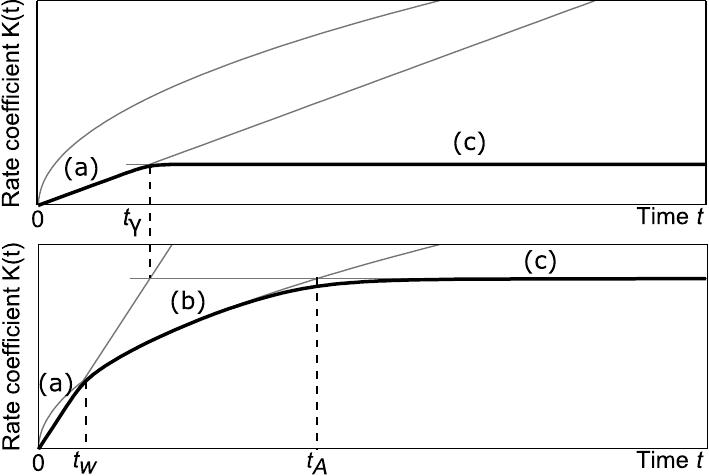}\hfill{}

\caption{\label{fig:evolution}Schematic evolution of the instantaneous rate
coefficient (thick black curve) showing the three regimes (a), (b)
and (c). Upper graph: moderate coupling ($t_{w}\gtrsim t_{A}$). Lower
graph: strong coupling ($t_{w}\ll t_{A}$). The grey curves correspond
to the limiting expressions (\ref{eq:rate1},\ref{eq:rate2},\ref{eq:rate3}).
Note that the parabolic curve associated with Eq. (\ref{eq:rate2})
is a universal upper limit of the rate coefficient.}
\end{figure}

This rate coefficient depends only on molecular physics parameters
$g$, $w$, $\Delta^{\prime}$, and $\gamma$. In fact, it can be
shown to derive from the time-dependent two-body theory, except for
the term proportional to $f(t)$ in the numerator of $K(t)$. We find
that this term can be neglected when we compare with numerical calculations
based on Eqs.~(\ref{eq:Eq1},\ref{eq:Eq2},\ref{eq:Eq3bis}). This
indicates that the term $f(t)$ should be discarded from Eq.~(\ref{eq:Shifted1}).
Depending on the relative strengths of the terms in the denominator
of $K(t)$, the rate coefficient for a fixed detuning goes through
three subsequent regimes illustrated in Fig.~\ref{fig:evolution}:
linear with time (a), square root of time (b) and constant with time
(c). 

\begin{eqnarray}
K(t) & = & \frac{2}{\hbar^{2}}w^{2}t\quad\quad\mbox{ for }t\ll t_{w}\label{eq:rate1}\\
K(t) & = & \frac{4}{\pi}\Big(\frac{h}{M}\Big)^{3/2}\sqrt{t}\quad\mbox{ for }t_{w}\ll t\ll t_{A}\label{eq:rate2}\\
K(t) & = & \frac{2}{\hbar}\mbox{Im}\frac{w^{2}}{\Delta-\Delta^{\prime}-i\gamma/2}\;\mbox{for }t_{A}\ll t\label{eq:rate3}\end{eqnarray}

where we define the two-body time scales\begin{eqnarray*}
t_{w} & = & \frac{1}{(\pi/2)^{2}}\left(\frac{h}{M}\right)^{3}\left(\frac{\hbar}{w}\right)^{4}\\
t_{A} & = & (\pi/2)^{2}\frac{1}{\hbar^{2}}\left(\frac{M}{h}\right)^{3}\frac{w^{4}}{(\Delta-\Delta^{\prime})^{2}+(\gamma/2)^{2}}\end{eqnarray*}
The first two regimes correspond to transient dynamics of the pairs,
while the third regime yields the rate coefficient of the time-independent
two-body theory. Note that the second regime might not exist for small
coupling $w$, when $t_{w}>t_{A}$. Then the first and third regime
are separated by the time scale $t_{\gamma}=\sqrt{t_{w}t_{A}}$.

In principle, the two transients regimes (a) and (b) always occur,
provided the coupling is large enough. However, they might often occur
at very small times, so that they do not affect the system significantly.
In other words, the time scales $t_{w}$ and $t_{A}$ might be much
smaller than the time scale for the evolution of the gas. We define
this time scale as the time $t_{\rho}$ needed to reduce the atomic
condensate population by a half. Obviously, this time scale depends
on the initial density $\rho_{0}$ of the gas, as can be seen from
Eq.~(\ref{eq:RateEquation}). Three situations are possible:

\begin{itemize}
\item $t_{\rho}\gg t_{A},t_{\gamma}$, in which case the evolution of the
gas is governed by the constant rate coefficient (\ref{eq:rate3})
\item $t_{w}\ll t_{\rho}\ll t_{A}$, in which case the evolution of the
gas is governed by the transient rate coefficient (\ref{eq:rate2})
\item $t_{\rho}\ll t_{w},t_{\gamma}$, in which case the evolution of the
gas is governed by the transient rate coefficient (\ref{eq:rate1})
\end{itemize}
\begin{figure}
\hfill{}\includegraphics[scale=1.8]{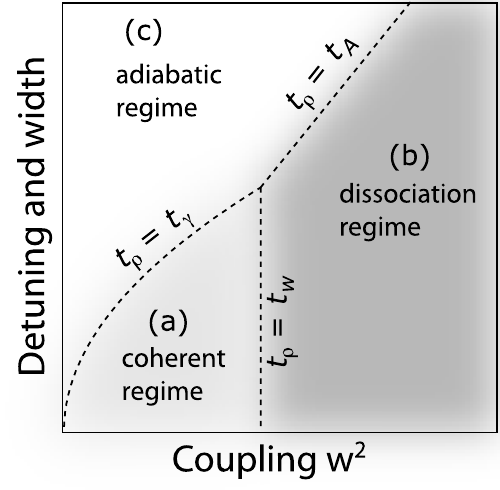}\hfill{}

\caption{\label{fig:regions} Different regimes of Feshbach resonance in a
condensate as a function of the detuning and width $\sqrt{(\Delta-\Delta^{\prime})^{2}+(\gamma/2)^{2}}$
and the coupling strength $w^{2}$, for a fixed density. The boundaries
are indicated by dashed lines and equalities of times scales defined
in the text.}
\end{figure}

We call these three situations the adiabatic regime, the dissociation
regime and the coherent regime, respectively. They are represented
in Fig.~\ref{fig:regions} for a fixed initial density, as a function
of the coupling strength $w^{2}$ and {}``width and detuning'' $\sqrt{(\Delta-\Delta^{\prime})^{2}+(\gamma/2)^{2}}$.
In the adiabatic regime, the molecular condensate $\Psi_{m}$ has
a very small population because it decays too fast or is too far from
resonance and it can be eliminated adiabatically. The system is then
described at all times by a single Gross-Pitaevski\u{\i} equation
such as (\ref{eq:GPeq}). In the dissociation regime, the molecular
condensate has also a small population, but now this is because it
is dissociated by the coupling into the pair continuum \cite{javanainen2002}.
The system can be described only in terms of the atomic condensate
and dissociated pairs. Interestingly, it is a universal regime in
the sense that all quantities depend only upon the mass of the atoms,
and not on the particular resonance used. In the coherent regime,
the molecular condensate can have a large population and dissociates
only into condensate atoms. This creates coherent oscillations between
the atomic and molecular condensates, which are described at all times
by two coupled Gross-Pitaevski\u{\i} equations. This regime corresponds
to the superchemistry described in \cite{heinzen2000}.

We stress that in all cases, the short-time dynamics of the atomic
condensate is described by the rate equation (\ref{eq:RateEquation})
with the time-dependent rate coefficient (\ref{eq:Rate}). This shows
that the loss of condensate atoms at short times goes like $t$, $t^{3/2}$
and $t^{2}$ for respectively the adiabatic, dissociation, and coherent
regimes. For longer times, the system is well described by the two
Gross-Pitaevski\u{\i} equations (\ref{eq:Shifted1}-\ref{eq:Shifted2})
containing the transient shift and broadening. However, higher-order
cumulants (quantum fluctuations) neglected in (\ref{eq:OriginalSet}),
as well as inelastic collisions \cite{yurovsky2003} between molecules,
dissociated pairs or condensate atoms play a role for long times (typically
$\sim$10 $\mu$s in the cases shown in Fig.~\ref{fig:ensemble}).

\section{Applications}

Experimentally, the adiabatic regime has been well explored with both
optical and magnetic Feshbach resonances \cite{cornish2000,mckenzie2002,theis2004,winkler2005}.
As explained above, the system is then described by a Gross-Pitaevski\u{\i}
equation where the scattering length appearing in the mean-field term
is changed by the resonance. In the case of an optical resonance,
the scattering length has an imaginary part because of losses from
the molecular state by spontaneous emission. These losses are described
by the rate equation (\ref{eq:RateEquation}) with the constant rate
coefficient (\ref{eq:rate3}).

\begin{figure}
\hfill{}\includegraphics[scale=0.26]{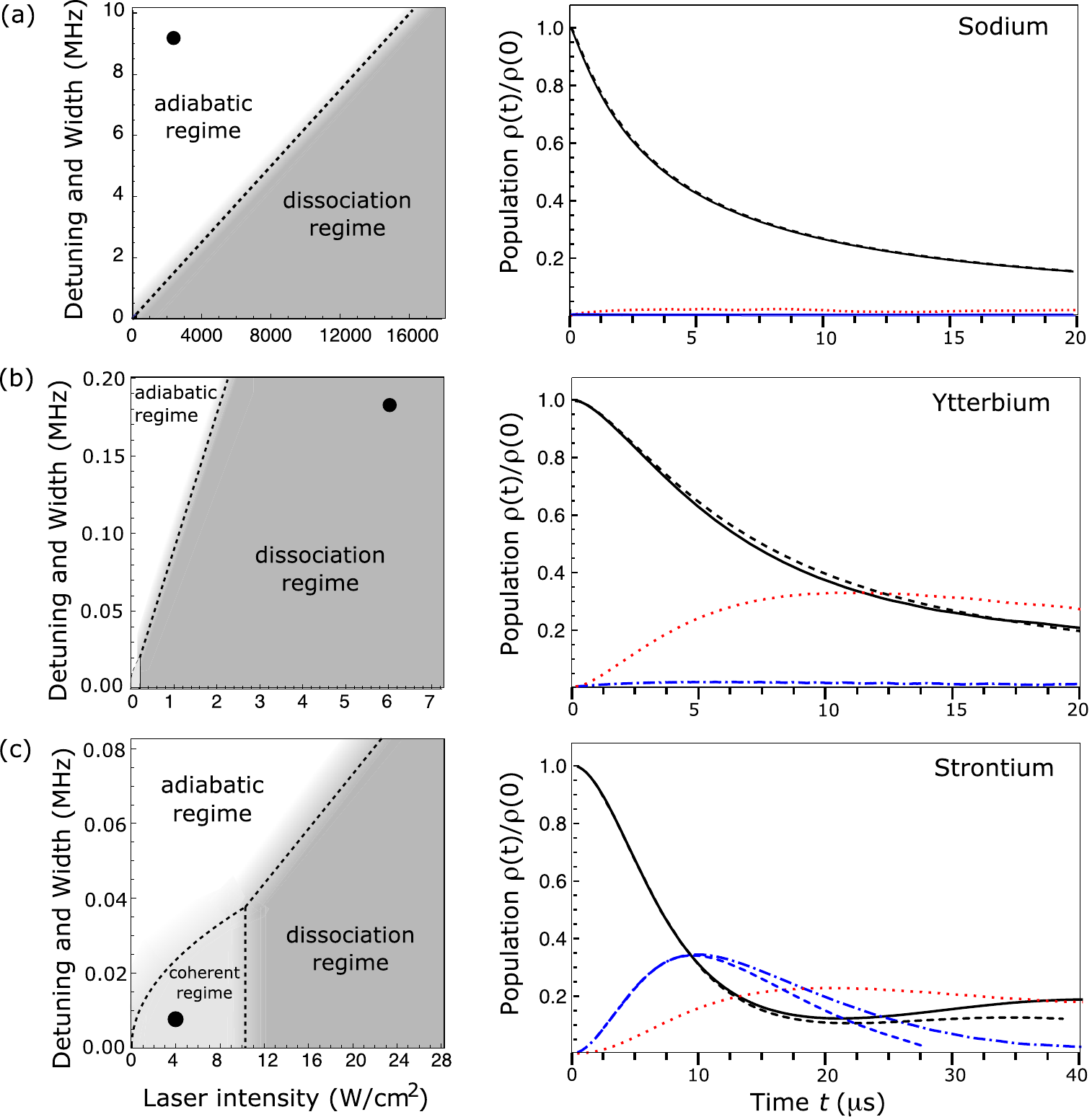}\hfill{}

\caption{\label{fig:ensemble}On-resonance photoassociation ($\Delta-\Delta^{\prime}=0$)
of a uniform atomic condensate of sodium (a), ytterbium (b) and strontium
(c), for typical transitions. In each row, the right panel the time
evolution of the populations, according to Eq. (\ref{eq:Eq1},\ref{eq:Eq2},\ref{eq:Eq3}),
based on the paramaters indicated by the black dot in the left panel,
which is a regime diagram similar to Fig. \ref{fig:regions}. Solid
curve: atomic condensate population; dotted curve: dissociated pair
population; dot-dashed curve: molecular population. The short-dashed
curve shows the atomic and molecular population following from Eqs.
(\ref{eq:Shifted1}-\ref{eq:Shifted2}).}

We used the following parameters: $w=1.255\cdot10^{-38}$ J$\cdot\mbox{m}^{3/2}$,
$\gamma/\hbar=18$ MHz, $w=3.850\cdot10^{-39}$ J$\cdot\mbox{m}^{3/2}$,
$\gamma/\hbar=364$ kHz for case (b), $w=6.347\cdot10^{-40}$ J$\cdot\mbox{m}^{3/2}$,
$\gamma/\hbar=15$ kHz for case (c). For all three cases, the initial
density is $\rho_{0}=6\cdot10^{14}\mbox{cm}^{-3}$.
\end{figure}

The dissociation regime has proved more difficult to observe. For
optical resonances, the losses from the molecular state usually confines
the system to the adiabatic regime, as can been seen from Fig.~\ref{fig:ensemble}.
To reach the dissociation regime, one needs to increase the coupling,
\emph{ie} the intensity of the laser. A high-intensity experiment
was performed at NIST with a sodium condensate \cite{mckenzie2002},
but the laser power was not sufficient to reach the dissociation regime.
Experiments in the dissociation regime were performed with magnetic
resonances by switching the magnetic field suddenly on or very close
to resonance \cite{donley2001}. They resulted in an explosion of
hot atoms, and thus were called {}``Bosenova'' experiments. The
hot atoms have been identified by several authors with the dissociated
pairs discussed above \cite{mackie2002,milstein2003}, although several
alternative theories exist \cite{duine2001,Santos2002,Saito2002,Adhikari2005}.
Quantitative comparison with any theory has not been completely satisfactory
so far \cite{Wuester2007}, suggesting that either some element is
missing from the theory or some experimental condition has been misunderstood.

To our knowledge, the coherent regime has not been observed. Here
we propose to observe the dissociation and coherent regimes by using
optical resonances in systems with narrow intercombination lines,
for which the molecular states are long-lived. Figure shows the cases
of ytterbium and strontium, for typical resonances. In the case of
ytterbium, it appears possible to reach the universal regime of dissociation
for laser intensities of about 3 W/$\mbox{cm}^{2}$. In the case of
strontium for similar intensities, one period of coherent oscillation
between the atomic and molecular condensates can be observed.

To illustrate some approximations made in the previous section, we
made numerical calculations based on equations (\ref{eq:Eq1},\ref{eq:Eq2},\ref{eq:Eq3})
and (\ref{eq:Shifted1},\ref{eq:Shifted2}) in the case of a uniform
system. Fig. \ref{fig:ensemble} shows the evolution of the populations
in the atomic, molecular condensate and dissociated pairs. One can
see that the short-time dynamics of equations (\ref{eq:Eq1},\ref{eq:Eq2},\ref{eq:Eq3})
is well described by the coupled Gross-Pitaevski\u{\i} equations
with the transient shift and broadening, Eqs. (\ref{eq:Shifted1}-\ref{eq:Shifted2}).

\begin{figure}
\hfill{}\includegraphics[scale=0.7]{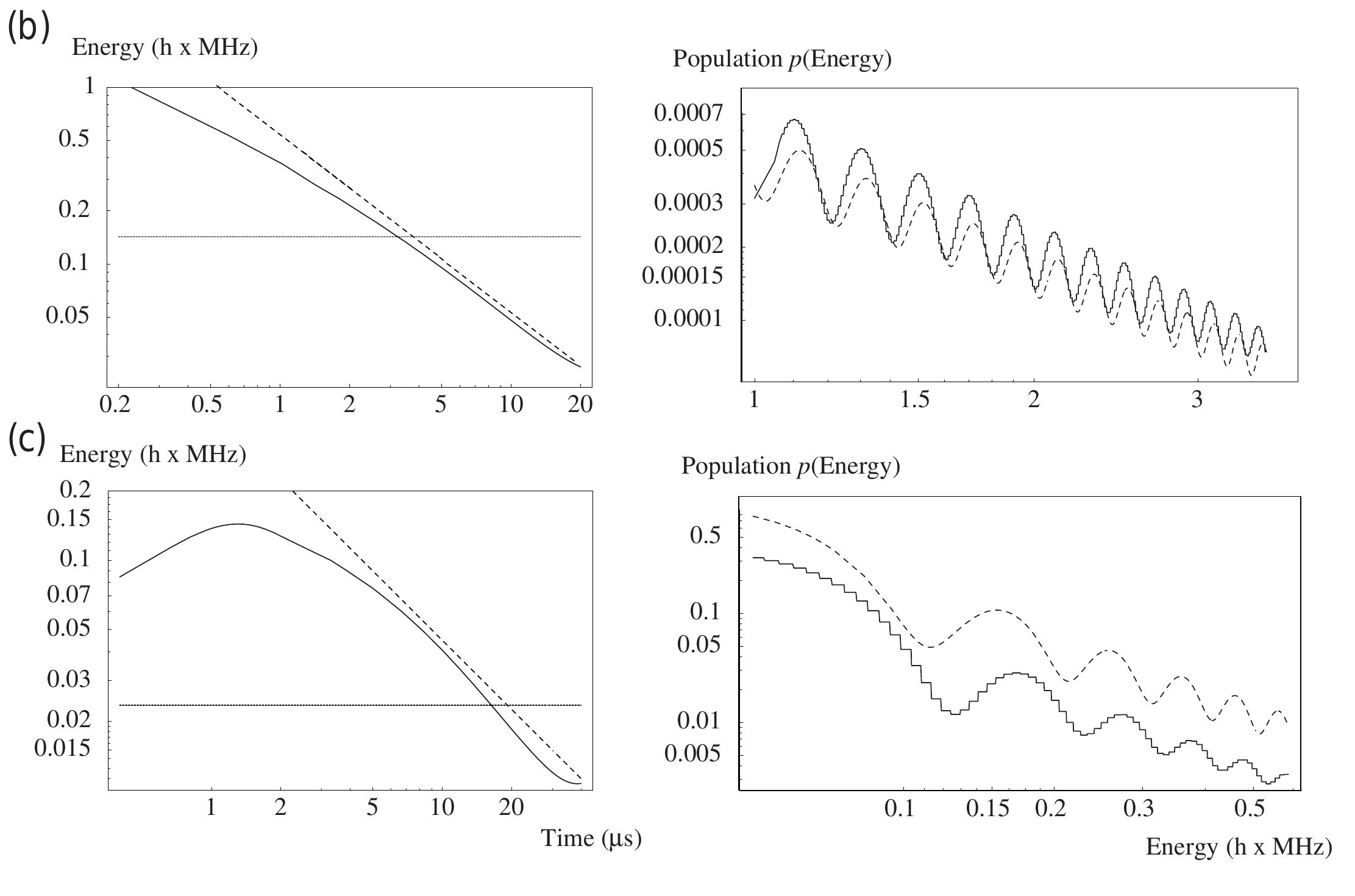}\hfill{}

\caption{\label{fig:Energy}Energy of the noncondensate atoms (dissociated
from the molecular state and not returning to the condensate) for
the cases (b) and (c) of Fig. \ref{fig:ensemble} (ytterbium and strontium).
Note that all graphs are in logarithmic scales. Left panels: average
energy as a function of time. For long times, the average energy decreases
as $1/t$ curves - the dashed lines are arbitrary $\propto1/t$ curves
shown for reference. Right panel: energy distribution at $t=5\,\mu$s
for case (b) and $t=10\,\mu$s for case (c). The dashed curves are
obtained from Eq.~(\ref{eq:Distribution}) . The jagged appearance
of the curves comes from the numerical discretisation of the pair
continuum.}
\end{figure}

Finally, we present in Fig.~\ref{fig:Energy} the energy distribution
of the dissociated pairs for the different systems. In the universal
regime, the dissociated pair distribution can be calculated analytically\begin{equation}
C_{\vec{k}}(\vec{x})=\Big[\frac{4}{k^{2}}\sqrt{i\frac{2ht}{M}}-\frac{4\pi i}{k^{3}}\mbox{Erf}\left(k\sqrt{-i\frac{\hbar t}{M}}\right)e^{-i\hbar k^{2}t/m}\Big]\Psi(\vec{x})^{2}\label{eq:Distribution}\end{equation}
where Erf is the error function. From this, one finds the energy distribution
$p(E_{k})\propto\frac{M}{4\pi\hbar^{2}}k\vert C_{k}\vert^{2}$. Figure
\ref{fig:Energy} shows the average energy of the dissociated pairs
as a function of time, and the final energy distribution. Although
Eq.~(\ref{eq:Distribution}) is expected to work only at short time
and in the universal regime, it gives a good qualitative prediction
of the energy distribution (up to a normalisation factor). It is worth
noting that when the laser is turned off quickly, although the number
of dissociated pairs is maintained, their energy might go down. We
find that the laser has to be turned off very fast to maintain the
final energy distribution. Therefore there is a range of switch-off
speed for which the number of pairs is maintained, while their energy
is lowered.

\section{Conclusion}

We showed that three limiting regimes are expected when coupling an
atomic condensate to a molecular condensate by an optical or magnetic
Feshbach resonance. The three regimes can be identified according
to the short-time dynamics of the system. This short-time dynamics
can be understood at the two-body level by the transient response
of the atom pairs, which is well described by a time-dependent shift
and broadening of the molecular state. This shift and broadening can
be incorporated into a set of coupled Gross-Pitaevski\u{\i} equations
to describe the short-time dynamics of the system. Finally, we suggested
different systems to observe the three regimes experimentally, in
particular the universal regime where atom pairs dissociate, and predicted
their short-time energy distribution.

\section{Appendix}

In this appendix, we give the derivation of Eqs.~(\ref{eq:Shifted1}-\ref{eq:Shifted2}).

\subsection*{Introduction of coupling constants}

We first express the pair cumulant $\Phi^{\prime}$ in terms of centre
of mass and relative motion by expanding it over the eigenstates $\varphi_{\vec{k}}(\vec{r})$
of the relative Hamiltonian\[
\Big(-\frac{\hbar^{2}\nabla_{r}^{2}}{M}+U(\vec{r})\Big)\varphi_{\vec{k}}(\vec{r})=E_{\vec{k}}\varphi_{\vec{k}}(\vec{r}),\]
where $E_{\vec{k}}=\hbar^{2}k^{2}/M$. One can write:\begin{equation}
\Phi^{\prime}(\vec{x},\vec{y},t)=\Phi^{\prime}(\vec{R},\vec{r},t)=\int\frac{d^{3}\vec{k}}{(2\pi)^{3}}C_{\vec{k}}(\vec{R},t)\varphi_{\vec{k}}(\vec{r}).\label{eq:DecompositionPhi}\end{equation}
For simplicity, only the continuum eigenstates $\varphi_{\vec{k}}(\vec{r})$
of the spectrum are taken into account (the bound states are supposed
to be far off-resonant) and we choose the following normalisation:
$\langle\varphi_{\vec{k}}\vert\varphi_{\vec{p}}\rangle=(2\pi)^{3}\delta(\vec{k}-\vec{p})$.

Inserting the decompositions (\ref{eq:DecompositionPhim}) and (\ref{eq:DecompositionPhi})
into the equations (\ref{eq:OriginalSet}), one obtains:\begin{eqnarray*}
i\hbar\frac{\partial\Psi}{\partial t}(\vec{x}) & = & H_{x}\Psi(\vec{x})+\tilde{g}(\vec{x})+\int\Psi^{*}(\vec{r}+\vec{x})\Big(U(\vec{r})\int\frac{d^{3}\vec{k}}{(2\pi)^{3}}C_{\vec{k}}(\frac{\vec{r}}{2}+\vec{x})\varphi_{\vec{k}}(\vec{r})+W(\vec{r})\Psi_{m}(\frac{\vec{r}}{2}+\vec{x})\varphi_{m}(\vec{r})\Big)d^{3}\vec{r}\\
i\hbar\frac{\partial}{\partial t}C_{\vec{k}}(\vec{R}) & = & (-\frac{\hbar^{2}\nabla_{R}^{2}}{4M}+E_{\vec{k}})C_{\vec{k}}(\vec{R})+\int\frac{d^{3}\vec{q}}{(2\pi)^{3}}v_{\vec{k}\vec{q}}(\vec{R})C_{\vec{q}}(\vec{R})+g_{\vec{k}}(\vec{R})+w_{\vec{k}}\Psi_{m}(\vec{R})\\
i\hbar\frac{\partial\Psi_{m}}{\partial t}(\vec{R}) & = & (-\frac{\hbar^{2}\nabla_{R}^{2}}{4M}+\Delta-i\gamma/2+v_{m}(\vec{R}))\Psi_{m}(\vec{R})+\tilde{w}(\vec{R})+\int\frac{d^{3}\vec{k}}{(2\pi)^{3}}w_{\vec{k}}^{*}C_{\vec{k}}(\vec{R})\end{eqnarray*}
where \begin{eqnarray*}
\tilde{g}(\vec{x}) & = & \int\Psi^{*}(\vec{r}+\vec{x})U(\vec{r})\Psi(\vec{x}+\frac{\vec{r}}{2})\Psi(\vec{x}-\frac{\vec{r}}{2})d^{3}\vec{r}\\
g_{\vec{k}}(\vec{R}) & = & \int\varphi_{\vec{k}}(\vec{r})U(\vec{r})\Psi(\vec{R}+\frac{\vec{r}}{2})\Psi(\vec{R}-\frac{\vec{r}}{2})d^{3}\vec{r}\\
\tilde{w}(\vec{R}) & = & \int\varphi_{m}(\vec{r})W(\vec{r})\Psi(\vec{R}+\frac{\vec{r}}{2})\Psi(\vec{R}-\frac{\vec{r}}{2})d^{3}\vec{r}\\
w_{\vec{k}} & = & \int\varphi_{\vec{k}}(\vec{r})W(\vec{r})\varphi_{m}(\vec{r})d^{3}\vec{r}\\
v_{m}(\vec{R}) & = & \int\varphi_{m}(\vec{r})\Big(V(\vec{R}+\frac{\vec{r}}{2})+V(\vec{R}-\frac{\vec{r}}{2})\Big)\varphi_{m}(\vec{r})d^{3}\vec{r}\\
v_{\vec{k}\vec{q}}(\vec{R}) & = & \int\varphi_{\vec{k}}(\vec{r})\Big(V(\vec{R}+\frac{\vec{r}}{2})+V(\vec{R}-\frac{\vec{r}}{2})\Big)\varphi_{\vec{q}}(\vec{r})d^{3}\vec{r}\end{eqnarray*}
are various matrix elements of the couplings $U$,$W$, and $V$ between
the states $\Psi$, $\Psi_{m}$ and $C_{\vec{k}}$.

The condensate wave function $\Psi$ and centre-of-mass function $\Psi_{m}$
and $C_{\vec{k}}$ extend over mesoscopic scales, typically up to
the micrometre range. On the other hand, quantities such as the interaction
potential $U$, $U_{m}$ and the molecular state $\varphi_{m}$ extend
over microscopic scales, on the nanometre scale. Therefore, we can
make the following approximations\begin{eqnarray*}
\tilde{g}(\vec{x}) & \approx & \Big(\int U(\vec{r})d^{3}\vec{r}\Big)\Psi^{*}(\vec{x})\Psi(\vec{x})^{2}\equiv\tilde{g}\Psi^{*}(\vec{x})\Psi(\vec{x})^{2}\\
g_{\vec{k}}(\vec{x}) & \approx & \Big(\int\varphi_{\vec{k}}(\vec{r})U(\vec{r})d^{3}\vec{r}\Big)\;\;\Psi(\vec{x})^{2}\equiv g_{\vec{k}}\Psi(\vec{x})^{2}\\
\tilde{w}(\vec{x}) & \approx & \Big(\int\varphi_{m}(\vec{r})W(\vec{r})d^{3}\vec{r}\Big)\;\;\Psi(\vec{x})^{2}\equiv\tilde{w}\Psi(\vec{x})^{2}\\
v_{m}(\vec{x}) & \approx & 2V(\vec{x})\end{eqnarray*}
which leads to\begin{eqnarray}
i\hbar\frac{\partial\Psi}{\partial t}(\vec{x}) & = & H_{x}\Psi(\vec{x})+\Psi^{*}(\vec{x})\Big(\tilde{g}\Psi(\vec{x})^{2}+\int\frac{d^{3}\vec{k}}{(2\pi)^{3}}g_{\vec{k}}C_{\vec{k}}(\vec{x})+w_{m}\Psi_{m}(\vec{x})\Big)\label{eq:Unren1}\\
i\hbar\frac{\partial}{\partial t}C_{\vec{k}}(\vec{x}) & = & \Big(-\frac{\hbar^{2}\nabla_{x}^{2}}{4M}+E_{\vec{k}}\Big)C_{\vec{k}}(\vec{x})+\int\frac{d^{3}\vec{q}}{(2\pi)^{3}}v_{\vec{k}\vec{q}}(\vec{x})C_{\vec{q}}(\vec{x})+g_{\vec{k}}\Psi(\vec{x})^{2}+w_{\vec{k}}\Psi_{m}(\vec{x})\label{eq:Unren2}\\
i\hbar\frac{\partial\Psi_{m}}{\partial t}(\vec{x}) & = & \Big(-\frac{\hbar^{2}\nabla_{x}^{2}}{4M}+2V(\vec{x})+\Delta-i\gamma/2\Big)\Psi_{m}(\vec{x})+\tilde{w}\Psi(\vec{x})^{2}+\int\frac{d^{3}\vec{k}}{(2\pi)^{3}}w_{\vec{k}}^{*}C_{\vec{k}}(\vec{x})\label{eq:Unren3}\end{eqnarray}
This is not the most convenient representation for this set of equations,
because in the case of a potential $U(\vec{r})$ which is strongly
repulsive at short distance, quantities such as $\tilde{g}$ might
diverge. Often, one treats this problem by replacing the actual potential
$U(\vec{r})$ by a contact pseudopotential $g\delta^{3}(\vec{r})$,
where $g$ is adjusted to some relevant finite value. This formally
replaces $\tilde{g}$ by $g$, but also introduces ultraviolent divergences,
due to the infinitely short range nature of the delta function. One
then has to renormalise the coupling constant $g$ using standard
techniques in field theory, so as to reproduce physical results. However,
in the present case, the whole set of equation is in fact well-behaved
with respect to the interaction $U$ and we do not need to resort
to such renormalisation techniques, but simply rewrite the equation
in another representation.

We decompose $C_{\vec{k}}$ into two contributions, the adiabatic
and dynamic ones:\[
C_{\vec{k}}(\vec{x})=C_{\vec{k}}^{ad}(\vec{x})+C_{\vec{k}}^{dyn}(\vec{x})\]
The adiabatic part is defined as the response to the time-dependent
source term $g_{\vec{k}}\Psi(\vec{x})^{2}+w_{\vec{k}}\Psi_{m}(\vec{x})$
in Eqs.~(\ref{eq:Unren2}), and is defined as the solution of the
system of equations: \begin{equation}
0=\Big(-\frac{\hbar^{2}\nabla_{x}^{2}}{4M}+E_{\vec{k}}\Big)C_{\vec{k}}^{ad}(\vec{x})+\int\frac{d^{3}\vec{q}}{(2\pi)^{3}}v_{\vec{k}\vec{q}}(\vec{x})C_{\vec{q}}^{ad}(\vec{x})+g_{\vec{k}}\Psi(\vec{x})^{2}+w_{\vec{k}}\Psi_{m}(\vec{x})\label{eq:AdiabaticEquation}\end{equation}
obtained by setting $\partial C_{\vec{k}}/\partial t=0$ in Eq.~(\ref{eq:Unren2}).
It is formally solved as\begin{equation}
C_{\vec{k}}^{ad}(x)=\int\!\! d^{3}\vec{y}\int\!\! d^{3}\vec{q}\; G_{\vec{k}\vec{q}}(\vec{y}-\vec{x})\Big(g_{\vec{q}}\Psi(\vec{y})^{2}+w_{\vec{q}}\Psi_{m}(\vec{y})\Big)\label{eq:ExactCkAdiab}\end{equation}
where $G_{\vec{k}\vec{q}}(\vec{y}-\vec{x})$ is a Green's function
associated with Eq.~(\ref{eq:AdiabaticEquation}). When introducing
$C_{\vec{k}}^{ad}$ in Eqs.~(\ref{eq:Unren1}-\ref{eq:Unren2}),
it appears only in integrals over momenta $\hbar\vec{q}$ where the
high-momentum contributions are the most significant. For sufficiently
high momenta, we can use a simplified version of (\ref{eq:ExactCkAdiab})\begin{equation}
C_{\vec{k}}^{ad}(\vec{x})=-\frac{g_{\vec{k}}\Psi(\vec{x})^{2}+w_{\vec{k}}\Psi_{m}(\vec{x})}{E_{\vec{k}}}.\label{eq:SimplifiedCkAdiab}\end{equation}
This amounts to neglecting the effects of the trap, \emph{i.e.} the
integral term and the Laplace operator in Eq.~(\ref{eq:AdiabaticEquation}).
Using this expression, we finally obtain\begin{eqnarray}
i\hbar\frac{\partial\Psi}{\partial t}(\vec{x}) & = & \Big(-\frac{\hbar^{2}\nabla_{x}^{2}}{2M}+V(\vec{x})+g\vert\Psi(\vec{x})\vert^{2}\Big)\Psi(\vec{x})+\Psi^{*}(x)\Big(w\Psi_{m}(\vec{x})+\int\frac{d^{3}\vec{k}}{(2\pi)^{3}}g_{\vec{k}}C_{\vec{k}}^{dyn}(\vec{x})\Big)\label{eq:Eq1}\\
i\hbar\frac{\partial\Psi_{m}}{\partial t}(\vec{x}) & = & \Big(-\frac{\hbar^{2}\nabla_{x}^{2}}{4M}+2V(\vec{x})+\Delta-\Delta^{\prime}-i\gamma/2\Big)\Psi_{m}(\vec{x})+w\Psi(\vec{x})^{2}+\int\frac{d^{3}\vec{k}}{(2\pi)^{3}}w_{\vec{k}}^{*}C_{\vec{k}}^{dyn}(\vec{x})\label{eq:Eq2}\end{eqnarray}
where\begin{eqnarray}
g & = & \tilde{g}-\int\!\!\frac{d^{3}\vec{k}}{(2\pi)^{3}}\frac{g_{\vec{k}}^{2}}{E_{\vec{k}}}\label{eq:Param1}\\
w & = & \tilde{w}-\int\!\!\frac{d^{3}\vec{k}}{(2\pi)^{3}}\frac{w_{\vec{k}}g_{\vec{k}}}{E_{\vec{k}}}\label{eq:Param2}\\
\Delta^{\prime} & = & \int\!\!\frac{d^{3}\vec{k}}{(2\pi)^{3}}\frac{w_{\vec{k}}^{2}}{E_{\vec{k}}}\label{eq:Param3}\end{eqnarray}
Unlike $\tilde{g}$ and $\tilde{w}$ , the coupling constants $g$
and $w$ are well-defined quantities. The constant $g$ corresponds
to the effective interaction strength between condensate atoms (one
can show that $g=g_{\vec{0}}=4\pi\hbar^{2}a/M$ where $a$ is the
s-wave scattering length associated to the potential $U$). The constant
$w$ is the effective coupling between condensate atoms and molecules,
and one can show that $w=w_{\vec{0}}$. The quantity $\Delta^{\prime}$
is a shift of the molecular energy due to the atomic-molecular coupling,
commonly referred to as {}``light shift'' in the case of an optical
resonance.

Equations (\ref{eq:Eq1}), (\ref{eq:Eq2}), and Eq.~(\ref{eq:ExactCkAdiab})
along with the equation\begin{equation}
i\hbar\frac{\partial}{\partial t}C_{\vec{k}}^{dyn}(\vec{x})=\Big(-\frac{\hbar^{2}\nabla_{x}^{2}}{4M}+E_{\vec{k}}\Big)C_{\vec{k}}^{dyn}(\vec{x})+\int\frac{d^{3}q}{(2\pi)^{3}}v_{\vec{k}\vec{q}}(\vec{x})C_{\vec{q}}^{dyn}(\vec{x})-i\hbar\frac{\partial}{\partial t}C_{\vec{k}}^{ad}(\vec{x})\label{eq:Eq3}\end{equation}
 form a closed set of equations. Often, $C_{\vec{k}}$ deviates significantly
from zero for small $\vec{k}$'s which lie in the Wigner's threshold
law regime \cite{mottmassey}. In this regime, one can make the approximation%
\footnote{This restricts the validity of our results to time scales larger than
$t_{C}=mr_{C}^{2}/\hbar$, where $r_{C}$ is on the order of the Condon
point radius for the molecular transition (\emph{ie} the typical size
of the molecule) or the van der Waals length associated to the interaction
potential $U$, whichever largest. In practice, these length scales
are very small and $t_{C}$ lies in the nanosecond range, which is
much smaller than all the other relevant time scales.%
} that $g_{\vec{k}}\approx g$ and $w_{\vec{k}}\approx w$. As a result,
the whole set of equations is determined by the three molecular parameters
$g$, $w$ and $\Delta^{\prime}$. These quantities can be measured
experimentally or calculated from the precise knowledge of the molecular
potentials $U$, $U_{m}$ and $W$.

\subsection*{Solution of the equations for an instantaneous coupling}

In general, the resolution of Eqs.~(\ref{eq:Eq1}), (\ref{eq:Eq2}),
and (\ref{eq:Eq3}) is involved and requires numerical calculation.
To proceed further, we apply a local density approximation to $C_{\vec{k}}^{dyn}$,
namely we neglect again the integral term and the laplacian in Eq.~(\ref{eq:Eq3}),
\begin{equation}
i\hbar\frac{\partial}{\partial t}C_{\vec{k}}^{dyn}(\vec{x})=E_{\vec{k}}C_{\vec{k}}^{dyn}(\vec{x})+\frac{i\hbar}{E_{\vec{k}}}\frac{\partial}{\partial t}\Big(g\Psi(\vec{x})^{2}+w\Psi_{m}(\vec{x})\Big)\label{eq:Eq3bis}\end{equation}
so that the spatial dependence of $C_{\vec{k}}^{dyn}$ comes only
from that of $\Psi$ and $\Psi_{m}$. Thus, we make some error regarding
the influence of the external trap $V$. This error should be small
for times much smaller than the typical oscillation times in the trap.
In the following, we will see that the interesting dynamics occurs
at the microsecond timescale, while trap oscillations are typically
around the millisecond timescale. This approximation is therefore
justified.

We are interested in the case where we start from a purely atomic
condensate and suddenly turn on the coupling $w$ to a molecular state.
The initial conditions are\begin{eqnarray}
\Psi(\vec{x},0) & = & \sqrt{\rho_{0}(\vec{x})}\label{eq:InitialConditions}\\
\Psi_{m}(\vec{x},0) & = & 0\nonumber \\
C_{k}^{dyn}(\vec{x},0) & = & 0\nonumber \end{eqnarray}
where $\rho_{0}(\vec{x})$ is an initial density profile. Integrating
Eq. (\ref{eq:Eq3bis}) gives\[
C_{\vec{k}}^{dyn}(\vec{x},t)=\int_{0}^{t}e^{\frac{i}{\hbar}E_{k}(\tau-t)}\frac{1}{E_{\vec{k}}}\frac{\partial}{\partial\tau}\Big(g\Psi(\vec{x},\tau)^{2}+w\Psi_{m}(\vec{x},\tau)\Big)d\tau.\]
This leads to

\begin{equation}
\int\frac{d^{3}\vec{k}}{(2\pi)^{3}}C_{\vec{k}}^{dyn}(\vec{x})=\int_{0}^{t}f(t-\tau)\frac{\partial}{\partial\tau}\Big(g\Psi(\vec{x},\tau)^{2}+w\Psi_{m}(\vec{x},\tau)\Big)d\tau\label{eq:IntCk}\end{equation}
with the complex function \begin{eqnarray*}
f(t-\tau) & = & \int\frac{d^{3}\vec{k}}{(2\pi)^{3}}\frac{e^{\frac{i}{\hbar}E_{k}(\tau-t+i\varepsilon)}}{E_{\vec{k}}}=\frac{1-i}{2\hbar}\Big(\frac{M}{2\pi\hbar}\Big)^{3/2}\frac{1}{\sqrt{t-\tau}}.\end{eqnarray*}
The $\varepsilon>0$ is introduced to maintain convergence of the
integral. It originates from the momentum dependence of $w_{\vec{k}}$,
which we ultimately neglected (which means that $\varepsilon$ is
set to zero in the end).

We now assume that the time integral in Eq. (\ref{eq:IntCk}) is dominated
by short-time contributions from $\frac{\partial}{\partial\tau}\Psi_{m}$,
which leads to the Ansatz\begin{equation}
\int_{0}^{t}f(t-\tau)\frac{\partial}{\partial\tau}\Big(g\Psi(\vec{x},\tau)^{2}+w\Psi_{m}(\vec{x},\tau)\Big)d\tau\approx\alpha f(t)w\Psi_{m}(\vec{x},t)\label{eq:Ansatz}\end{equation}

where $\alpha$ is a numerical factor to be determined. Inserting
this result in Eqs. (\ref{eq:Eq1}-\ref{eq:Eq2}), we finally get
the closed set of equations (\ref{eq:Shifted1}-\ref{eq:Shifted1}).

To determine $\alpha$, we consider the case where the transient shift
and broadening is dominant in Eq. (\ref{eq:Shifted2}), which gives\[
\Psi_{m}(\vec{x},t)=\frac{-\Psi(\vec{x},t)^{2}}{w\alpha f(t)}.\]

Inserting this expression in Eq. (\ref{eq:Ansatz}), we find $\alpha=\pi/2$.
Interestingly, the same reasoning can be applied to systems of reduced
dimensionality, which can be created with an external potential $V(\vec{x})$
strongly confining the atoms in one or two directions \cite{olshanii1998,petrov2000a}.
In this case, the motion of the atoms is frozen in the confined directions.
This renormalises the coupling parameters $g$, $w$ and $\Delta^{\prime}$
\cite{naidon2006b}, and the integration over momenta is reduced to
2 or 1 dimensions in all expressions. For a 2D-like system, we find\begin{eqnarray*}
f_{2D}(t) & = & \frac{1}{4\pi}\frac{M}{\hbar^{2}}e^{\frac{i}{\hbar}E_{0}t}\Gamma(0,\frac{i}{\hbar}E_{0}t)\sim-\frac{1}{4\pi}\frac{M}{\hbar^{2}}\log(E_{0}t/\hbar)\,\mbox{ for }\, t\ll\hbar/E_{0}\\
\alpha_{2D} & = & 1\end{eqnarray*}
where $\Gamma$ is the generalised gamma function, and $E_{0}$ is
the zero-point energy due to the confinement. For a 1D-like system,
however, one finds $\alpha_{1D}=0$, which indicates that the Ansatz
is invalid in this case.

\bibliographystyle{apsrev}
\bibliography{/home/pascal/Redaction/biblio,/home/pascal/Redaction/biblio_extra,/home/pascal/Redaction/pascal,/home/pascal/Redaction/Paper06b/bibliopaper,biblio,biblio_extra,/Users/pascal/Documents/Travail/Redaction/Paper06c/bibliopaper}

\end{document}